\documentclass[a4paper,11pt]{article}
\usepackage{setspace}
\usepackage{epsfig}
\usepackage{delarray}
\usepackage{amsmath, amssymb, amsfonts}
\usepackage{bm}
\usepackage{graphicx}
\usepackage{dcolumn}
\usepackage{float}

\begin{document}
\title{On Negativity and Quantum Fisher Information of an Open and Noisy System in the Steady State}
\maketitle
\author{Azmi Ali Altintas}\\
\small{Faculty of Engineering and Architecture, Okan University, Istanbul, Turkey}
\footnote{altintas.azmiali@gmail.com}

\begin{abstract}
In this work, we study the  quantum Fisher information (QFI) per particle of an open (particles can enter and leave the system) and  dissipative (far from thermodynamical equilibrium) steady state system of two qubits in noisy channels. We concentrate on two noisy channels these are dephasing and non-dephasing channels. We will show that under certain conditions QFI per particle is slightly greater than 1 for both systems. This means that both systems can be slightly entangled.  
\end{abstract}
\section{Introduction}
In quantum mechanics, it is hard to measure observables, instead we try to estimate them. Estimating a parameter is subject of information theory. Because of that quantum mechanics gets some tools from estimation theory to guess the parameters.
Also we can design some experiments to estimate parameters, for example Mach-Zender interferometer. It is used to determine the relative phase shift between two collimated beams. It is well known that entanglement can increase the sensitivity of interferometer. Quantum Fisher information (QFI)  characterizes the sensitivity of a quantum system with respect to the changes of a parameter of the system. It can be taken as a multipartite entanglement criteria\cite{PezzeSmerzi2009PRL,Hyllus2012PRA}: If the mean quantum Fisher information per particle of a state exceeds the so called \textit{shot-noise limit} i.e. the ultimate limit that separable states can provide, then the state is multipartite entangled.
 \textit{Shot-noise limit} is $\Delta \theta \equiv \frac{1}{\sqrt{N}}$, where N is the number of particles\cite{Xiong201010}.  Only for N=2 case,  any entangled state can be made useful by local operations \cite{Hyllus2010PRA}. It is also shown that GHZ states provide the largest sensitivity, achieving the fundamental, so called Heisenberg limit \cite{Lloyd2004Science}. Mean QFI determines the phase sensitivity of  state with respect to SU(2) rotations. Recently the quantum Fisher information has been further studied both theoretically and experimentally[6-26].
 
Usually the natural systems are open and noisy. If a quantum system interacts with environment it is thought as the quantum system is in a noisy channel. It is obvious that noise decreases the entanglement of quantum system. If the system is open the decrease can be balanced by determining a reset mechanism.  With the help of reset mechanism an entangled steady state can be established. The reset mechanism replaces randomly system particles with particles from the environment in some standard, sufficiently pure, single-particle state \cite{Briegel2006PRA}. Reset mechanism, itself, can not produce entanglement. To create entanglement, the fresh particles must interact with the system.  Reset mechanism requires particle exchange from the environment, this brings that the system must be open. Hartmann et.al shown that for both gas type and strongly coupled quantum systems the effect of decoherence can be vanished with the help of reset mechanism \cite{Hartmann2007NJP}.  
    
In this work, we study the quantum Fisher information per particle of open and dissipative noisy system of two qubits with reset mechanism. We will concentrate on two different types of noise. Firstly we take dephasing channel\cite{Altintas}, then secondly we look at non-dephasing channel and we try to find states whose mean QFI is greater than 1. We examine the effect of reset mechanism for dephasing case, and for non-dephasing case we will determine a temperature dependent parameter ``s" then we will examine effect of ``s" on QFI.    
\section{QFI of open noisy system in a steady state}
Quantum Fisher information of a given quantum system can be written from \cite{Hyllus2010PRA} as;
\begin{equation}
F(\rho,J_{\overrightarrow{n}})=\sum_{i\neq j}\frac{2(p_i-p_j)^2}{p_i+p_j}{|\langle i |J_{\overrightarrow{n}}| j \rangle|}^2=\overrightarrow{n}\textbf{C}\overrightarrow{n}^T.
\end{equation}
Here $p_i$ and $| j \rangle$ are the eigenvalue and eigenvector of state $\rho$ respectively. Also $\overrightarrow{n}$  is a normalized three dimensional vector and $J_{\overrightarrow{n}}=\sum_{\alpha=x,y,z} \frac{1}{2}n_\alpha \sigma_\alpha$, the angular momentum operator in $\overrightarrow{n} $ direction. $\sigma_\alpha $ are Pauli  matrices. Also, $p_i+p_j=0$ terms are not included to summation.  After some calculations the matrix elements of the symmetric matrix \textbf{C} can be found as;
\begin{equation}
\textbf{$C_{kl}$}=\sum_{i\neq j}\frac{(p_i-p_j)^2}{p_i+p_j}[\langle i | J_{k}|j \rangle\langle j | J_{l}|i \rangle+
\langle i | J_{l}|j \rangle\langle j | J_{k}|i \rangle]
\end{equation}
If $\rho$ is a pure state the equation 2 is written as
\begin{equation}
\textbf{$C_{kl}$}=2\langle J_kJ_l+J_lJ_k\rangle-4\langle J_k\rangle\langle J_l\rangle,
\end{equation}
and the QFI is also expressed as $F(\rho,J_{\overrightarrow{n}})=4(\Delta J_{\overrightarrow{n}} )^2$.
The mean QFI is found as in \cite{Ma2011PRA}
\begin{equation}
\overline{F}_{max}=\frac{F_{max}}{N}=\frac{\lambda_{max}}{N}
\end{equation}
here $\lambda_{max}$ is maximum eigenvalue of matrix C and N is the number of particles. Also $\lambda_{max}$ is the maximum value of QFI. 
It has recently been shown that,  the QFI for separable states is \cite{PezzeSmerzi2009PRL}
\begin{equation}
\overline{F}_{max}\leq 1 
\end{equation} 
and 
for general  states the mean QFI of the system is
\begin{equation}
\overline{F}_{max}\leq N
\end{equation} 
where the bound $\overline{F}_{max}=N$ can be saturated by maximally entangled states.

Now, we define an open quantum system with reset mechanism in a noisy channel. The total master equation which defines the quantum system is given by \cite{Briegel2006PRA}
\begin{equation}\label{master}
\dot{\rho}=-i[H,\rho]+\textit{$L_{noise}\rho$}+r\sum_{i=1}^N(|\chi_i\rangle_i\langle\chi_i|tr_i\rho-\rho)
\end{equation}
The master equation is in form of Lindbald equation. The first term in the right hand side of eq. (\ref{master}) is just about total Hamiltonian of the quantum system, the second term describes the noisy channel and the third term describes the reset mechanism and N is the number of particle.
\begin{eqnarray}
L_{noise}\rho&=&\sum_{i=1}^N-\frac{B}{2}(1-s)[\sigma_+^i\sigma_{-}^i\rho+\rho\sigma_+^i\sigma_{-}^i-2\sigma_{-}^i\rho\sigma_{+}^i]\\\nonumber
&-&\frac{B}{2}s[\sigma_{-}^i\sigma_{+}^i\rho+\rho\sigma_{-}^i\sigma_{+}^i-2\sigma_{+}^i\rho\sigma_{-}^i]
-\frac{2C-B}{4}[\rho-\sigma_{z}^i\rho\sigma_{z}^i]
\end{eqnarray}
Here B is inversion and C is polarization parameter. Also s is a temperature dependent parameter which is 
\begin{equation}
s=(e^{\omega\beta}+1)^{-1}\in[0,1]
\end{equation}
Since $\beta=\frac{1}{T}$, when $T\rightarrow\infty$ s will be 0.5.

\subsection{Dephasing Case}

In this case we choose dephasing channel as a noisy channel. The Hamiltonian of two qubit steady state is
\begin{equation}
 H=g\overrightarrow{\sigma}^{(1)}_z\overrightarrow{\sigma}^{(2)}_z,
 \end{equation}
here $g\geq 0$ is the coupling strength.
Since the noise is a dephasing channel the second term in equation (\ref{master}) is
\begin{equation}
\textit{$L_{noise}\rho$}=\frac{\gamma}{2}\sum_{i=1,2}(\sigma_{z}^{(i)}\rho\sigma_{z}^{(i)}-\rho)
\end{equation}
here $\gamma$ is strength of decoherence which is a positive real number. One can get this expression by choosing $B=0$ and $C=\gamma$ in equation (8). The expression of reset mechanism is written by taking $N=2$ as
\begin{equation}
\textit{$L_{reset}\rho$}=r\sum_{i=1,2}(|\chi_i\rangle_i\langle\chi_i|tr_i\rho-\rho).
\end{equation}

Since the reset state should be able to produce entanglement from the resulting product state,  the reset state must be depend on the Hamiltonian. For example for our Hamiltonian we can not choose the reset state as $|\chi_i\rangle_i=|0\rangle$, since the state does not create any entanglement. Then our two qubit master equation becomes
\begin{equation}
\dot{\rho}=-i[H,\rho]+\frac{\gamma}{2}\sum_{i=1,2}(\sigma_{z}^{(i)}\rho\sigma_{z}^{(i)}-\rho)+r\sum_{i=1,2}(|+\rangle_i\langle +|tr_i\rho-\rho).
\end{equation}
When $r=0$, the noise  destroys the entanglement. For $r\rightarrow \infty$ case Hamiltonian and noise parts are neglected and the reset part injects
fresh particles to the system. Thus the entanglement between two qubits will be zero eventually.
 $\rho$ is the density state and it can be expressed as matrix form. In our case it is $4\mbox{x}4$ matrix. The matrix is written from \cite{Briegel2006PRA}, 
 \begin{equation}
\rho_{11}=\rho_{22}=\rho_{33}=\rho_{44}=\frac{1}{4},
\end{equation}
 \begin{equation}
\rho_{14}=\rho_{23}=\rho_{32}=\rho_{41}=\frac{r^2(r+\gamma/2)}{4(r+\gamma)[2g^2+(r+\gamma/2)(r+\gamma)])},
\end{equation}

\begin{eqnarray}
\rho_{12}&=&\rho_{13}=\rho_{42}=\rho_{43}=\frac{r(-ig+r+\gamma/2)}{4[2g^2+(r+\gamma/2)(r+\gamma)]},\\
\rho_{21}&=&\rho_{24}=\rho_{31}=\rho_{34}=\frac{r(ig+r+\gamma/2)}{4[2g^2+(r+\gamma/2)(r+\gamma)]}.
\end{eqnarray}
Here r, $\gamma$ and g are reset, decoherence and coupling strength parameters respectively and they are real parameters.

Now by using $\rho$ matrix in equation (2) we find  QFI per particle of the system depending on parameters r, $\gamma$ and g. 

 To understand the behavior of the  QFI per particle first we fixed the noise parameter to $\gamma=0.5.$ 
 
\begin{center}
\begin{figure}[H]
\includegraphics[width=0.5\textwidth]{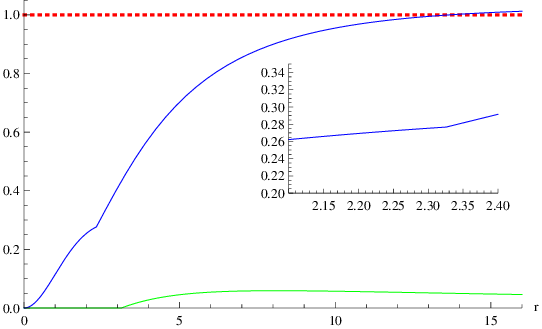}
\caption{( QFI per particle (blue) and negativity (green) vs reset with $\gamma$ = 0.5. Red dotted line represents the shot noise limit. Inset shows the critical point where the optimal direction changes.) } \label{fig:Fisher}
\end{figure}
\end{center}
As one can see from the figure at  $r=0$ the QFI of the system is 0 as expected. When r is at 14  QFI per particle has a value as 1.00226. Well known entanglement criteria is  negativity and it can take values between 0 and 1. For our entangled state negativity is 0.0496243. It means that the chosen state is weakly entangled. 
When negativity is 0 the state is separable, when the negativity equals to 1 the state is maximally entangled. 

For the figure (1) ($g=2.5$ and $\gamma=0.5$ case) the optimal direction $\overrightarrow{n}^o=\overrightarrow{n}_1$ when $r \leq 2.3$. For $r>2.3$ the optimal direction  $\overrightarrow{n}^o=\overrightarrow{n}_2sin(\frac{\pi}{2})+\overrightarrow{n}_3cos(\frac{\pi}{2})$. Here $\overrightarrow{n}_1$ is unit vector in x axis. $\overrightarrow{n}_2\; \mbox{and}\; \overrightarrow{n}_3$ are unit vectors in y and z directions respectively. 

\subsection{Non-Dephasing Case}

In this case, the noisy channel is a non-dephasing channel. In equation \ref{master} the hamiltonian contains free and interaction terms.
\begin{equation}
H_{free}=\frac{\omega}{2}\sum_i^N\sigma_{z}^i\;\mbox{and}\;H_{int}=g\sigma_x^1\sigma_x^2.
\end{equation}
Also in noise part we take $C=\frac{B}{2}$ and $\omega=B$. 
In the light of above, one can solve the master equation and the elements of density matrix will be
\begin{equation}
\rho_{11}=\frac{B^2s^2w^2+(B+2r)((B+r)g^2+B^2(B+2r)s^2)}
{(B+r)((B+r)\omega^2+(B+2r)(4g^2+(B+r)(B+2r)))},
\end{equation}
\begin{equation}
\rho_{14}=\rho_{41}^\ast\frac{g(2sB-B-r)(-i(B+2r)+\omega)}{((B+r)\omega^2+(B+2r)(4g^2+(B+r)(B+2r)))},
\end{equation}
\begin{equation}
\rho_{22}=\rho_{33}=\frac{(B+2r)((B+r)g^2-B^2(B+2r)s^2+B(B+r)(B+2r)s)-Bs(sB-B-r)\omega^2}
{(B+r)((B+r)\omega^2+(B+2r)(4g^2+(B+r)(B+2r)))},
\end{equation}
\begin{equation}
\rho_{44}=\frac{(-Bs+B+r)\omega^2+(B+2r)(B^2(B+2r)s^2-2B(B+r)(B+2r)s+(B+r)(g^2+(B+r)(B+2r))}
{(B+r)((B+r)\omega^2+(B+2r)(4g^2+(B+r)(B+2r)))},
\end{equation}
The other terms of density matrix is are 0.

Now by using the density matrix in equation (2) we find the mean QFI of the system. To understand the effect of temperature we take s as a free parameter, it can have values between 0 and 0.5. We fix the reset parameter as $r=10$ and we choose as $\omega=1$

\begin{center}
\begin{figure}[H]
\includegraphics[width=0.5\textwidth]{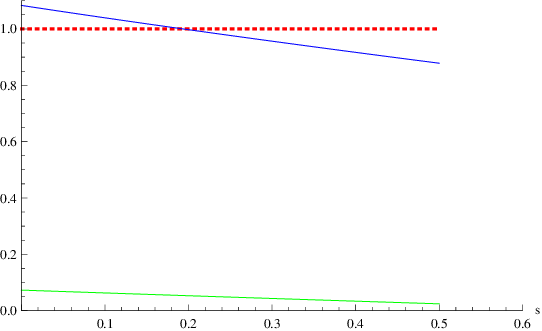}
\caption{(QFI per particle (blue) and negativity (green) vs s with r = 10. Red dotted line represents the shot noise limit.)} \label{fig:Fisher2}
\end{figure}
\end{center}

It can be easily seen that when s is greater than 0.193 mean QFI is smaller than 1 although negativity is still greater than 0.  This means that QFI does not recognize all the entangled states. 	

\section{Conclusion}

 We have studied the quantum Fisher information of a noisy open quantum system of two qubits which is in a steady state. In the system we use an interaction Hamiltonian in a noisy channel and a reset mechanism. By solving master equation we have defined density state in matrix form and with the help of equation 2 we describe mean QFI of the system depending on reset and noise parameters. 

We have shown the change of  QFI per particle depending on reset parameter. In figure 1 we have chosen $\gamma$ as $0.5$ and have observed that at $r=14$ QFI per particle is greater than 1. The negativity of such state is greater than 0. It means that the state is entangled\cite{Altintas}. 

Also we take non-dephasing noisy channel as second example. At that case we take reset parameter constant as $r=10$ and we define a temperature dependent parameter s. The change of QFI per particle depending on parameter s is investigated and figure 2 shows that until s be 0.193 QFI per particle is greater than 1. After that value of s QFI per particle is smaller than 1 although negativity is greater than 0. 

These two example show us that QFI does not recognize all entangled states. Despite the QFI is entanglement witness it can not be taken as entanglement measure. 

There are some recent works on q-deformation and quantum information theory\cite{Gavrilik, AltintasQIP}, in the light of these works as a further work one can calculate the QFI of q deformed version of the states that we introduce this study.

\end{document}